\documentclass[12pt]{article}

\usepackage{amssymb}
\if@twoside  m
\else
\fi
\begin{document}

 \newcommand{\nid}{\noindent}
 \newcommand{\ie}{{\em i.e.}}
 \newcommand{\eg}{{\em e.g.}}
 \newcommand{\cf}{{\em cf. }}
 \newcommand{\Cf}{{\em Cf. }}
 \newcommand{\esa}{{\em e.s.a. }}
 \newcommand{\etc}{{\em etc. }}
 \newcommand{\Iff}{{\em iff $\:$}}
 \newcommand{\sgn}{{\rm sgn\,}}
 \newcommand{\lfrac}{\lbrack\!\lbrack\!}
 \newcommand{\rfrac}{\!\rbrack\!\rbrack}
 \newcommand{\QED}{\mbox{\rule[-1.5pt]{6pt}{10pt}}}
 \newcommand{\restr}{\vert\hskip -5.5pt \phantom{\vert}^{\scriptscriptstyle
            \backslash}}
 \newcommand{\rhs}{{\em rhs }}
 \newcommand{\Tr}{{\rm Tr\,}}
 \newcommand{\re}{{\rm Re\,}}
 \newcommand{\im}{{\rm Im\,}}
 \newcommand{\C}{\mathbb{C}}
 \newcommand{\R}{\mathbb{R}}
 \newcommand{\Z}{\mathbb{Z}}
 \newcommand{\AAA}{{\cal A}}
 \newcommand{\DD}{{\cal D}}
 \newcommand{\EE}{{\cal E}}
 \newcommand{\LL}{{\cal L}}
 \newcommand{\OO}{{\cal O}}
 \newcommand{\eps}{\varepsilon}
 \newtheorem{rem99}{Remark~99$\,$-$\!\!$}
 \newtheorem{claim}{Claim}[section]
 \newtheorem{theorem}[claim]{Theorem}
 \newtheorem{lemma}[claim]{Lemma}
 \newtheorem{proposition}[claim]{Proposition}
 \newtheorem{corollary}[claim]{Corollary}
 \newtheorem{remark}[claim]{Remark}
 \newtheorem{remarks}[claim]{Remarks}

\title{Some properties of the one-dimensional \\ generalized point 
interactions (a torso)}
\date{}
\author{Pavel Exner$^{1,2}$ {\em and$\:$} Harald Grosse$^{3,4}$}
\maketitle
\begin{quote}
1 {\em Nuclear Physics Institute, Academy of Sciences, CZ--25068
\v Re\v z near \\ \phantom{1 }Prague, Czechia} \\
2 {\em Doppler Institute for Mathematical Physics, Czech Technical
\\ \phantom{1 }University, B\v{r}ehov\'{a}~7, CZ--11519 Prague,
Czechia}
\\
3 {\em Institut f\"ur Theoretische Physik, Universit\"at Wien,
Boltzmanngasse 5, \\ \phantom{1 }A--1090 Vienna, Austria} \\
4 {\em E.Schr\"odinger Institute for Mathematical Physics,
Boltzmanngasse 9, \\ \phantom{1 }A--1090 Vienna, Austria}
\end{quote}

\noindent {\small This text is a part of an unfinished project 
which deals with the generalized point interaction (GPI) in one 
dimension. We employ two natural parametrizations, which are 
known but have not attracted much attention, to express the 
resolvent of the GPI Hamiltonian as well as its spectral and 
scattering properties. It is also shown that the GPI yields one 
of the simplest models in which a non-trivial Berry phase is 
exhibited. Furthermore, the generalized Kronig-Penney model 
corresponding to the GPI is discussed. We show that there are 
three different types of the high-energy behaviour for the 
corresponding band spectrum.}

\section{Introduction}

Many projects have a complicated history and some never make it 
to a paper; most of us will find examples on our desks. The 
present text was conceived in the fall of 1993 as a part of a 
larger study. For various reasons the final result never 
materialized and the draft could be easily put into the bin. If 
we do not do that it is because it contains some results on 
generalized point interactions in one dimension which may be of 
an independent interest. We reproduce the text as it was written 
six years ago, updating the references and adding in places an 
occasional ``remark~99" to reflect the current state of affairs. 

\begin{center}
* \quad * \quad *
\end{center}

\noindent The intuitively attractive idea of describing 
interaction of quantum particles with sharply localized objects 
by $\,\delta$-shaped potentials was introduced in the early days 
of quantum mechanics \cite{Fe,KP}. However, it was only in the 
beginning of the sixties when Berezin and Faddeev \cite{BF} 
suggested how such formal Schr\"odinger operators can be 
constructed as mathematically well defined objects, namely as 
self-adjoint extensions of a symmetric operator which coincides 
with the free Hamiltonian outside the support of the 
interaction. 

Two decades later point interactions became an object of a
systematic and extensive study which was summarized in the
monograph \cite{AGHH}. At the same time, numerous generalizations
has appeared with contact-type interactions on configuration
spaces of a non-trivial geometric structure, for relativistic
Hamiltonians, with applications to the perturbation theory of
embedded eigenvalues, \ie, decay and resonance models \etc -- a
list of references can be found, \eg, in \cite{AESS,E1}.

\begin{rem99} We complete the list of references with several
new items -- see \cite{ADK}--\cite{RT} -- without striving for
completeness. The most exhaustive bibliography to the date can be
found in the forthcoming monograph by Albeverio and Kurasov
\cite{AK}.
\end{rem99}

Somehow unnoticed remained in this developments remarkable
properties of one-dimensional point interactions. Recall that --
as long as one stays within the standard quantum mechanical
setting -- the self-adjoint extension construction works for
dimensions $\,d\leq 3\,$, because otherwise a restriction of the
Laplacian to functions which vanish in the vicinity of a fixed
point yields an \esa operator. There is a substantial difference,
however, between $\,d=2,3\,$ on one side and one-dimensional
systems on the other coming from the fact that a one-point
restriction of the one-dimensional Laplacian leads to deficiency
indices $\,(2,2)\,$, and therefore to a four-parameter family of
self-adjoint extensions.

The generalized point interaction (GPI) was introduced by \v Seba
\cite{Se1}. Until recently the only particular case of it
different from the standard point (or $\,\delta\,$) interaction
which was discussed was the so-called {\em
$\,\delta'$-interaction} \cite{AGHH}, and even this did not
attract the attention of physicists because of the lack of a
reasonable physical model. Recall that in distinction to
$\,\delta\,$, the $\,\delta'$-interaction cannot be approximated
by a family of Schr\"odinger operators with squeezed potentials;
in this sense the name is misleading because $\,\delta'\,$ is {\em
not} an elementary dipole.

Instead, there are other approximations. The first of them was
found by \v Seba \cite{Se2} who demonstrated that $\,\delta'$ is a
limit of a suitable sequence of rank-one perturbations to the free
Hamiltonian. Alternatively, one can use scaled Schr\"odinger
operators but with velocity-dependent potentials. On a formal
level, this was suggested for a two-parameter family of extensions
in \cite{Se1}.  Recently, another approximation of this type
(using non-selfadjoint Schr\"odinger operators) for a
four-parameter class of GPI's including the
$\,\delta'$-interaction has been suggested in \cite{Ca}, and a
similar procedure has been proposed for another three-parameter
class "almost disjoint" with the former one \cite{CH}. Still
another possibility -- physically a very exciting one -- is based
on the observation that the scattering properties of $\,\delta'\,$
can be reproduced in a fixed energy interval by a suitable
many-loop graph; in this sense $\,\delta'$-interactions appears to
a paradigm for geometric scatterers \cite{AEL1}.

\begin{rem99} Another geometric scatterer with similar properties
is a sphere with two leads -- see \cite{Kis} and \cite{ETV} --
although the scattering in this case is more complicated. The
above claim about the impossibility of approximation by a family
of Schr\"odinger operators with squeezed potentials is not quite
correct -- it was shown recently [CS] that one can do that with
potentials scaled in a {\em nonlinear} way. A rigorous nature of
this approximation, however, remains to be clarified.
\end{rem99}

Moreover, it is known that the $\,\delta'\,$ modification of the
Kronig-Penney model exhibits gaps whose widths are growing at
large energies - \cf \cite[Sec.III.3]{AGHH}. If a homogeneous
electric field is added, this leads to rather interesting spectral
properties \cite{E2} which are quite unlike those of the
conventional Wannier-Stark ladder \cite{Ne}. In particular, such
Hamiltonians appear to have empty absolutely continuous part of
the spectrum, and the rest is likely to depend substantially on
the slope of the linear potential: if the latter has a rational
value in suitable units, the spectrum is pure point and nowhere
dense, while in an irrational case it covers the whole real line.
The proof of Ref.\cite{E2} cannot be adapted for the $\,\delta\,$
Wannier-Stark ladders whose spectral properties remain still an
open problem. This gives a strong motivation for study the
analogous problem for the general GPI including the the cases
``intermediary'' between the $\,\delta\,$ and $\,\delta'\,$.

\begin{rem99} A part of the original plan was to extend the
result about the absence of absolutely continuous spectrum to
other $\,\delta'$-type GPI's. The question is still there, but in
the course of time other aspects of the $\,\delta'$ Wannier-Stark
problem appeared to be more appealing. In particular, the above
claim about the essential spectrum (formulated as a conjecture in
\cite{AEL1, E2}) has been proved, and moreover, the spectrum has
been shown to be pure point for a ``large" set of irrational
slopes \cite{ADE}.
\end{rem99}

Another motivation comes from the search for simple models
exhibiting a nontrivial geometric phase. Recently its existence
has been demonstrated for a quantum particle on an interval with a
family of boundary conditions coupling the endpoints \cite{GK}.
Since the occurrence of eigenvalue crossings is essential for the
effect, it cannot be achieved with a standard Schr\"odinger
operator on {\em line} having a potential which is limit-point at
both $\,\pm\infty\,$, because the corresponding spectrum is
simple. Neither can any of the standard point interactions be
used, since they have at most one eigenvalue. Unlike the
$\,\delta\,$ and $\,\delta'\,$, the one-center GPI has in general
two eigenvalues which {\em do} cross at finite values of the
parameters, and therefore it might yield the simplest example of a
system with a nontrivial geometric phase. We shall show that this
is indeed the case.

\begin{rem99} We intended also to look into the behaviour of the
continuous spectrum when the GPI parameters change. This 
appeared to be less urgent after the paper [SA] was published 
where analogous question was discussed in a more general 
context. Notice, however, that the Berry phase of the example 
given in Section~3 is independent of the parameter loop size 
exhibiting thus the ``homeopathic" behaviour investigated in the 
framework of another model in [AB]. 
\end{rem99}

It is not our intention to write an exhaustive study which would
constitute another chapter of \cite{AGHH}; we want to concentrate
primarily on the two above mentioned physically interesting problems.
However, since several authors have addressed already the question,
each of them using his own notation, a general introduction and
mutual comparison is needed.

Let us review briefly the contents of the paper. In the next
section we first introduce two natural parametrizations of the GPI
and compare them to those existing in the literature. Then we
derive an explicit expression for the resolvent kernel and use it
to discuss spectral properties of the one-center GPI Hamiltonian,
in particular, its eigenvalues and eigenfunctions. We also find
the corresponding scattering matrix and show how it behaves at low
and high energies.

In Section~3 we present the mentioned example of a geometric phase
arising when the coupling-constant vector makes a loop in the
parameter space.

In Section~4 we study equidistant arrays of GPI's. We show that
the spectrum of the generalized Kronig-Penney model has always
infinitely many gaps, however, their behaviour depends
substantially on the parameters of the GPI. In addition to the
$\,\delta\,$ and $\,\delta'$-type situations, where the
gap-to-band width ratio is decreasing and growing, respectively,
we specify a class of the GPI's for which this ratio is
asymptotically constant with respect to the band number.

\setcounter{equation}{0}
\section{The one-center generalized point interaction}

\subsection{Boundary conditions}

Without loss of generality, we may assume that the mass is
$\,m=1/2\,$ and the interaction is supported by the point $\,x=0\,$.
The standard construction starts from the restriction of the free
Hamiltonian $\,H_0:=-d^2/dx^2\,$ with $\,D(H_0):=H^{2,2}(\R)\,$ to
the subspace $\,\DD:=\{\,f\in D(H_0):\: f(0)=f'(0)=0\,\}\,$, which is a
symmetric operator with the deficiency indices $\,(2,2)\,$.

The most straightforward way to get the corresponding family of
self-adjoint extensions is to use the von Neumann theory as \v
Seba did in his pioneering paper \cite{Se1}, see also \cite{CH}.
If the operators under consideration are ordinary differential
ones, however, it is more suitable to use boundary conditions; the
drawback is that they usually become singular for some values of
the parameters. It is easy to write a general four-parameter
family of boundary conditions. Various choices have been used in
\cite{Ca,CH,GeK,Ku,Se1}; below we shall present their comparison.

Here we propose two other sets of boundary conditions which seem us
to be natural for the problem under consideration. The first of them
is the following
\begin{eqnarray} \label{Greek bc}
f'(0+)-f'(0-) &=& \frac{\alpha}{2}\,(f(0+)\!+\!f(0-))\,+\,
\frac{\gamma}{2}\, (f'(0+)\!+\!f'(0-))\,, \nonumber \\ \\
f(0+)-f(0-) &=& -\,\frac{\bar\gamma}{2}\,(f(0+)\!+\!f(0-))\,+\,
\frac{\beta}{2}\, (f'(0+)\!+\!f'(0-)) \nonumber
\end{eqnarray}
with $\,\alpha,\beta\in\R\,$ and $\,\gamma\in\C\,$. For brevity,
denote $\,\AAA:= \left({\scriptsize \begin{array}{cc} \alpha & \gamma
\\ -\bar\gamma & \beta \end{array}}\right)\,$. This form of the
boundary conditions reduces easily to the standard cases: for
$\,\beta=\gamma=0\,$ we get the $\,\delta$-interaction with the
``coupling constant'' $\,\alpha\,$, while $\,\alpha=\gamma=0\,$
yields the $\,\delta'$-interaction of strength $\,\beta\,$. The
family (\ref{Greek bc}) describes almost all self-adjoint
extensions of $\,H_0\restr\,\DD\,$, with the exception of the
four-point set in the parameter space referring to the situations
where the Dirichlet or Neumann conditions are imposed from {\em
both} sides of the point $\,x=0\;$ (see also
Remark~\ref{symmetries}a below).  The other family of boundary
condition we shall use is
\begin{equation} \label{lc bc}
f'(0+)= af(0+)+cf(0-)\,, \quad\, -f'(0-)= \bar cf(0+)+bf(0-)
\end{equation}
with $\,a,b\in\R\,$ and $\,c\in\C\;$; its advantage is that it allows
to describe in a simple way the subset of the parameter space where
the conditions decouple and the two halflines become independent.
\begin{proposition} \label{bc}
Any of the above boundary conditions define a self-adjoint
extension to $\,H_0\restr\,\DD\,$. The conditions (\ref{Greek bc})
decouple separating the motion of the left and the right halflines
\Iff $\;c=0\,$, which is further equivalent to
\begin{equation} \label{decoupling}
\det\AAA=4\quad {\rm and}\quad\; \im\gamma=0\,.
\end{equation}
The correspondence between the boundary conditions is given by the
relations
\begin{equation} \label{lc-Greek}
\left( \begin{array}{cc} a & c \\ \bar c & b \end{array} \right)
\,=\, \frac{1}{4\beta}\: \left( \begin{array}{cc}
4+\det\AAA+4\re\gamma & -4+\det\AAA-4i\im\gamma \\
-4+\det\AAA+4i\im\gamma & 4+\det\AAA-4\re\gamma \end{array}
\right)\,,
\end{equation}
\begin{equation} \label{Greek-lc}
\left( \begin{array}{cc} \alpha & \gamma \\ -\bar\gamma & \beta
\end{array} \right) \,=\, \frac{4}{a+b-2\re c} \left(
\begin{array}{cc} ab-|c|^2 & {1\over 2}(a-b)-i\im c \\ -{1\over
2}(a-b)-i\im c & 1 \end{array} \right)\,,
\end{equation}
where $\,\det\AAA= \alpha\beta+|\gamma|^2=\, 4\,
\frac{a+b+2\re c}{a+b-2\re c}\,$, provided the denominators are
non-zero.
\end{proposition}
\begin{remarks} \label{HPA+uc}
   \begin{description} {\rm
   \item{\em (a)} The conditions (\ref{lc bc}) are in fact a particular
case (for $\,E=0\,$) of those used in \cite{E1}. More exactly, they
are related by the natural isomorphism $\,U:\: Uf= {\scriptsize
\left( \begin{array}{c} f_+ \\ f_- \end{array}\right)}\,$ between
$\,L^2(\R)\,$ and $\,L^2(\R^+)\oplus L^2(\R^+)\,$, where
$\,f_{\pm}(x):= f(\pm x)\,\restr\,\R^+\;$; the opposite sign of the
derivative in the second condition is due to the change of the
orientation of the negative halfline.
   \item{\em (b)} This shows, at the same time, that the GPI on line
is unitarily equivalent to the non-trivial (\ie, $\,s$-wave) part
of a ``two-channel'' point interaction in $\,\R^3\;$; in the
decoupled case we have in each channel just the point interaction
of the strength $\,\alpha/4\pi\,$ and $\,\beta/4\pi\,$,
respectively.
   \item{\em (c)} It is clear from (\ref{lc-Greek}) that the
conditions (\ref{lc bc}) make no sense if $\,\beta=0\,$. In this case
one can use a reformulation, namely
\begin{equation} \label{uc bc}
f(0+)= Af'(0+)-Cf'(0-)\,, \quad\, f'(0-)= \bar Cf(0+)-Bf(0-)
\end{equation}
with $\,A,B\in\R\,$ and $\,C\in\C\,$, where
$$
\left( \begin{array}{cc} A & -C \\ \bar C & -B \end{array}
\right) = \frac{1}{ab\!+\!|c|^2}\,\left( \begin{array}{cc} b & -c
\\ \bar c & -a \end{array} \right)\,,\;
\left( \begin{array}{cc} a & c \\ \bar c & b \end{array} \right) =
\frac{1}{AB\!-\!|C|^2}\, \left( \begin{array}{cc} B & -C \\ -\bar
C & A \end{array} \right) $$
provided again the denominators are
non-zero. The conditions (\ref{uc bc}) decouple clearly \Iff
$\;C=0\,$.
   }\end{description}
\end{remarks}

Let us further comment on relations between (\ref{Greek bc}),
(\ref{lc bc}) and the other known parametrizations of self-adjoint
extensions of the operator $\,H_0\restr\,\DD\,$:
   \begin{description}
   \item{\em (i)} an ``almost general'' set of boundary conditions
\begin{equation} \label{Kurasov}
f(0+)= \omega \tilde af(0-)+\omega \tilde bf'(0-)\,, \quad\, f'(0+)=
\omega \tilde cf(0-)+\omega \tilde df'(0-)\,,
\end{equation}
where $\,|\omega|=1\,$ and $\,\tilde a,\tilde b,\tilde c,\tilde
d\,$ are real numbers such that $\,\tilde a\tilde d\!-\!\tilde
b\tilde c=1\,$, have been used in \cite{CH,GeK,Ku,Se1}, sometimes
without the factor $\,\omega\,$ which can be removed by a unitary
transformation -- \cf Remark~\ref{symmetries}a below. They are
related to (\ref{Greek bc}), (\ref{lc bc}) by
\begin{equation} \label{lc-Kurasov}
a={\tilde d\over\tilde b}\,, \quad  b={\tilde a\over\tilde b}\,,
\quad c=-{\omega\over\tilde b}
\end{equation}
and
\begin{equation} \label{Greek-Kurasov}
\alpha={4\tilde c\over \tilde a+\tilde d+2\re\omega}\,, \quad
\beta={4\tilde b\over \tilde a+\tilde d+2\re\omega}\,, \quad
\gamma=2\,{\tilde d-\tilde a+2i\im\omega\over \tilde
a+\tilde d+2\re\omega}\,.
\end{equation}
This covers the $\,\delta\,$ and $\,\delta'$-interactions (for
$\,\omega=1,\: \tilde a=\tilde d,\: \tilde b=0\,$ with
$\,\alpha:=\tilde c\,$, and $\,\omega=1,\: \tilde a=\tilde d,\:
\tilde c=0\,$ with $\,\beta:=\tilde b\,$, respectively), while the
decoupled case is not included,
   \item{\em (ii)} in \cite{Ca}, the boundary conditions (\ref{lc
bc}) have been used, however, the parameters have been written in the
form
\begin{equation} \label{lc-Carreau}
a=\rho_c\!+\beta_c\,\quad b=\rho_c\!+\alpha_c\,, \quad c=-\rho_c\,
e^{-i\theta_c}
\end{equation}
with $\,\alpha_c,\beta_c\in\R\,,\; \rho_c\geq 0\,$ and $\,\theta_c\in
[0,2\pi)\,$. The relation to (\ref{Greek bc}) is
\begin{eqnarray} \label{Greek-Carreau}
\alpha &\!=\!& 4\,{\alpha_c\beta_c+\rho_c(\alpha_c+\beta_c)\over
\alpha_c+\beta_c+4\rho_c \cos^2\left({1\over 2}\theta_c\right)}\,,
\quad \beta \,=\, {4\over \alpha_c+\beta_c+4\rho_c
\cos^2\left({1\over 2}\theta_c\right)} \nonumber \\ \\
\gamma &\!=\!& 2\,{\beta_c-\alpha_c-2i\rho_c\sin\theta_c \over
\alpha_c+\beta_c+4\rho_c \cos^2\left({1\over 2}\theta_c\right)}\,.
\nonumber
\end{eqnarray}
The corresponding boundary conditions again do not cover the case
$\,\beta=0\,$ including the $\,\delta$-interaction. On the other
hand, $\,\delta'\,$ corresponds to
$\,\alpha_c=\beta_c=\theta_c=0\,$ and the coupling constant
$\,\beta:=\rho_c^{-1}\,$, and the decoupled case to
$\,\rho_c=0\,$,
   \item{\em (iii)} the following two-parameter family was studied
in \cite{Se1}
\begin{equation} \label{Seba}
f(0+)= -\gamma_s f(0-)-\delta_s f'(0-)\,, \quad f'(0+)= -\beta_s
f(0-)-\alpha_s f'(0-)\,,
\end{equation}
where the parameters involved are real numbers such that
$\,\alpha_s+\gamma_s=-2\,$ and $\,\alpha_s\gamma_s
-\beta_s\delta_s=1\,$. The parameters of (\ref{Greek bc}) and
(\ref{lc bc}) are then given by
\begin{equation} \label{Greek-Seba}
\alpha_s={(\gamma_s+1)^2\over\delta_s}\,, \quad \beta=-\delta_s\,,
\quad \gamma= \gamma_s+1
\end{equation}
and
\begin{equation} \label{lc-Seba}
a=-\, {\gamma_s+2\over \delta_s}\,, \quad
b={\gamma_s\over\delta_s}\,, \quad c={1\over\delta_s}\,.
\end{equation}
This yields the $\,\delta'$-interaction with $\,\gamma_s=-1\,$ and
$\,\beta:= -\delta_s\,$, while neither $\,\delta\,$ nor the
decoupled case make sense here,
   \item{\em (iv)} the three-parameter family
\begin{equation} \label{CH}
f(0+)=e^{-z}f(0-)\,, \quad rf(0+)+f'(0-)=e^{\bar z}(rf(0-)+f'(0-))
\end{equation}
has been used in \cite{CH}; it is disjoint with the previous one with
the exception of the free case. We have
\begin{equation} \label{Greek-CH}
\alpha= {4r(e^{2\re z}\!-1)\over |1+e^z|^2}\,, \quad \beta=0\,, \quad
\gamma= 2\, {e^{\bar z}\!-1 \over e^{\bar z}\!+1}\,.
\end{equation}
Hence this parametrization is suitable for the extensions with
$\,\beta=0\,$ which are covered neither by (\ref{lc bc}) nor by
(\ref{Seba}). We can, of course, use the modification of
Remark~\ref{HPA+uc}c for which the parameters are
\begin{equation} \label{uc-CH}
A= {1\over r(e^{2\re z}\!-1)}\,, \quad B= {e^{2\re z}\over r(e^{2\re
z}\!-1)}\,, \quad C= {e^{\bar z}\over r(e^{2\re z}\!-1)}\,.
\end{equation}
It should be stressed that the Chernoff-Hughes parametrization
does {\em not} cover the ``pure'' $\,\delta$-interaction either,
with the exception of the free case ($\,\alpha=0\,$). The
particular choice $\,r=0\,$ and $\,z\in\R\,$ has been considered
in \cite{Ku}; it corre\-sponds to the ``off-diagonal'' interaction
with $\,\alpha= \beta=0\,$ and $\,\gamma\in\R\,$.
   \end{description}

\begin{remark} \label{confusion}
{\rm There has been some confusion concerning the GPI's in recent
physical literature. Apart from a nonsensical proposal critized by
the authors of \cite{AGHH} in \cite{AGH}, there is a note
\cite{Gr} aiming at correction of the same mistake. The author has
arrived at the just mentioned ``off-diagonal'' conditions together
with the standard $\,\delta$-interaction ones, however, he
proposed also a generalization to ``higher derivatives of the
$\,\delta\,$ function'', failing to realize that such conditions
cannot yield a self-adjoint operator for the (second-order)
Schr\"odinger equation.}
\end{remark}

\begin{remarks} \label{symmetries}
   \begin{description} {\rm
   \item{\em (a)} The relations (\ref{CH}) show that the boundary
conditions (\ref{Greek bc}) and (\ref{Kurasov}) have a``hidden
degeneracy'', namely the zero values of all coefficients describe
together with the free Hamiltonian $\,H_0\,$ also the
one-parameter family of extensions which are unitarily equivalent
to $\,H_0\,$ by the operators $\,U_{\omega}:\, (U_{\omega}f)(x)=
(\Theta(-x)+ \omega\Theta(x))f(x)\;$; they could be dubbed
``quasifree''.
   \item{\em (b)} Since the time-reversal operator is represented by
complex conjugation on $\,L^2(\R)\,$, the extensions {\em
invariant w.r.t. the time reversal} are those with real
coefficients in the corresponding boundary conditions, \ie,
$\,\gamma,\, c,\, C\in\R\,$ in (\ref{Greek bc}), (\ref{lc bc}) and
(\ref{uc bc}), respectively; this includes both $\,\delta\,$ and
$\,\delta'$-interactions, as well as the decoupled case. For the
other boundary conditions mentioned, this requires $\,\omega=\pm
1\,$ for (\ref{Kurasov}) where, of course, the sign can be
absorbed into the coefficients; $\,\theta_c=0,\pi\,$ for the
parametrization (\ref{lc-Carreau}) and $\,z\in\R\,$ for
(\ref{CH}); the extensions given by (\ref{Seba}) are time-reversal
invariant.
   \item{\em (c)} Notice also that the operator $\,U_{\omega}\,$ of
{\em (a)} produces in general one-parameter families of unitarily
equivalent (and therefore isospectral) extensions corresponding to
fixed $\,a,\,b\,$ and $\,|c|\;$; among each family, just the
operators with $\,c=\pm |c|\,$ are time-reversal invariant.
   \item{\em (d)} In the same way, one can ask about extensions
{\em invariant w.r.t. the space reflection.} Since the boundary
values satisfy $\,(Rf)(0\pm)=f(0\mp)\,$ and
$\,(Rf)'(0\pm)=-f'(0\mp)\,$ for $\,R:\: (Rf)(x)=f(-x)\,$, we see
that this requires $\,\gamma=0\;$; hence every space-reflection
invariant extension is at the same time invariant w.r.t. the time
reversal. In the other parametrizations mentioned, the condition
is equivalent to $\,a=b\,$ and $\,c\in\R\,$ for (\ref{lc bc}), or
$\,\alpha_c=\beta_c,\: \theta_c=0,\pi\,$ for (\ref{lc-Carreau}),
and to $\,\tilde a=\tilde d,\: \omega=\pm 1\,$ for
(\ref{Kurasov}). No extension given by (\ref{Seba}) is
space-reflection invariant, while the class specified by
(\ref{CH}) has a trivial -- quasifree in the sense of {\em (a)} --
intersection with the space-reflection invariant extensions. In
particular, the $\,\delta\,$ and $\,\delta'$-interactions are
space reflection invariant; for the decoupled case this is true
\Iff $\;a=b\,$.
   }\end{description}
\end{remarks}

\subsection{The resolvent}

For the sake of brevity, we shall use the symbol $\,\AAA\,$ for a
general point in the parameter space referring to the appropriate
choice of the coefficients described above; the corresponding
self-adjoint extension of $\,H_0\restr\,\DD\,$ will be denoted
$\,H_{\AAA}\,$.

To analyse spectral properties of these operators, we need to know
the corresponding resolvent. It is, of course, an integral operator,
so we have to find the corresponding kernel.  We denote
conventionally $\,k:=\sqrt z\,$ with the cut along the positive real
axis.
\begin{proposition} \label{resolvent}
The resolvent kernel of $\,H_{\AAA}\,$ for $\,\AAA:= \left(
{\scriptsize \begin{array}{cc} \alpha & \gamma \\ -\bar\gamma &
\beta \end{array}} \right)\,$ is
\begin{eqnarray} \label{Greek kernel}
\lefteqn{G_{\AAA}(x,x';k)\,=\, {1\over k}\, \left(\Theta(x)
\Theta(x') e^{ikx_>}\sin kx_<\,-\,\Theta(-x)\Theta(-x') e^{ikx_<}
\sin kx_> \right)} \nonumber \\
&& +\,{\beta\over 2}\, F_{\AAA}(k)^{-1}\, \bigl\lbrace
\Theta(x)\Theta(x')\, (4+\det\AAA-4\re\gamma-
4ik\beta) e^{ik(x+x')} \nonumber  \phantom{AAAAAAA} \\
&& +\, \Theta(-x)\Theta(-x')\,(4+\det\AAA+4\re\gamma- 4ik\beta)
e^{-ik(x+x')} \\
&& +\, \Theta(x)\Theta(-x')\, (4-\det\AAA+4\im\gamma) e^{ik(x-x')}
\nonumber \\
&& +\,\Theta(-x)\Theta(x')\, (4-\det\AAA-4\im\gamma) e^{-ik(x-x')}
\bigr\rbrace\,, \nonumber
\end{eqnarray}
where $\,F_{\AAA}(k):= (\alpha\beta\!
+|\gamma|^2\!-2ik\beta) (2-ik\beta)- |\gamma|^2\,$, the
symbols $\,x_>,\, x_<\,$ mean the maximum and minimum of $\,x,\,x'\,$,
respectively, and $\,\Theta\,$ is the Heaviside function. In the
parametrization (\ref{lc bc}), it expresses as
\begin{eqnarray} \label{lc kernel}
\lefteqn{G_{\AAA}(x,x';k)\,=\, {1\over k}\, \left(\Theta(x)
\Theta(x') e^{ikx_>}\sin kx_<\,-\,\Theta(-x)\Theta(-x') e^{ikx_<}
\sin kx_> \right)} \nonumber \\
&& +\,D_{\AAA}(k)^{-1}\, \bigl\lbrace \Theta(x)\Theta(x')\,
(b\!-\!ik) e^{ik(x+x')}+\, \Theta(-x)\Theta(-x')\, (a\!-\!ik)
e^{-ik(x+x')}  \phantom{AAA} \\
&& -\, \Theta(x)\Theta(-x')\,c\, e^{ik(x-x')}-\,
\Theta(-x)\Theta(x')\, \bar c\, e^{-ik(x-x')}
\bigr\rbrace \nonumber
\end{eqnarray}
with $\,D_{\AAA}:= (a\!-\!ik)(b\!-\!ik)-|c|^2\,$.
\end{proposition}
\vspace{1mm}

\nid {\em Proof:} Using the Krein-formula argument from
Proposition~2.1 of \cite{E1} together with the unitary equivalence
of Remark~\ref{HPA+uc}a, we obtain the latter formula; the former
then follows from (\ref{lc-Greek}). \quad \QED \vspace{3mm}

In the particular case of $\,\delta$-interaction, $\,\AAA=
\left({\scriptsize \begin{array}{cc} \alpha & 0 \\ 0 & 0
\end{array}} \right)\,$, one can use fact that $\,{\beta \over
2}\,F_{\AAA}(k)^{-1}\to\, {1\over 4}\,(\alpha-2ik)^{-1}\;$ as
$\,\beta,\gamma\to 0\,$ together with the identity
$$
{1\over \alpha-2ik}\,=\, {i\over 2k}\,-\, {2k\alpha\over
2k+i\alpha}\, \left( {i\over 2k}\right)^2
$$
to check that the resolvent kernel reduces to the standard expression
$$
G_{\alpha}(x,x';k)\,=\, {i\over 2k}\, e^{ik|x-x'|}- \,
{2k\alpha\over 2k+i\alpha}\, \left( {i\over 2k}\right)^2 e^{ik|x|}
e^{ik|x'|}
$$
(\cf \cite{AGHH}, Chap.I.3). On the other hand, using the identities
$$ -\,{i\over 2k}\,+\,{i+k\beta\over k(2-ik\beta)}\,=\,
-\,\left({i\over 2\pi} \right)^2 {2\beta k^2\over 2-ik\beta}\,=\,
\,{i\over 2k}\,-\, {i\over k(2-ik\beta)}\,, $$
we check easily that for $\,\alpha=\gamma=0\,$ we arrive back at
the standard $\,\delta'$-expression
$$
G_{\beta}(x,x';k)\,=\, {i\over 2k}\, e^{ik|x-x'|}- \,
{2\beta k^2\over 2-ik\beta}\, \tilde G(x)\tilde G(x')\,,
$$
where $\,\tilde G(x):= {i\over 2k}\, e^{ikx} \sgn x\;$ (\cf
\cite{AGHH}, Chap.I.4). A similar simplification can be obtained in
the decoupled case which is not surprising, of course, because the
formula (\ref{lc kernel}) was constructed starting from the
decoupled resolvent \cite[Proposition~2.1]{E1}.

\subsection{Spectral properties}

Since the GPI under consideration represents a finite-rank
perturbation to (the resolvent of) $\;H_0\,$, the essential
spectrum is preserved, $\,\sigma_{ess}(H_{\AAA})=\R^+\,$.
Moreover, using the explicit form of the resolvent given above, it
is easy to check that there is no singularly continuous spectrum
(\cf \cite[Thm.XIII.9]{RS}), so $\,\sigma_{ac}(H_{\AAA})=\R^+\,$
and the only non-trivial effect the perturbation may produce are
eigenvalues of $\,H_{\AAA}\,$ without accumulation points (at most
two in $\,(-\infty,0)\,$).

We know from \cite{E1} that if a potential is added to the GPI,
the resolvent kernel may have a singularity at a general point of
the complex $\,k$-plane. In the present case, however, the
singularities are confined to the imaginary axis only, hence it is
useful to the quantity
$$
\kappa\,:=\,-ik\,.
$$
The spectral condition $\,D_{\AAA}(k)=0\,$ is then solved by
$\,k_{\pm}\,$ corresponding to
\begin{equation} \label{lc singularities}
\kappa_{\pm}\,=\, -\,{1\over 2}(a+b)\,\mp\,{1\over 2}
\sqrt{(a-b)^2+4|c|^2}\,,
\end{equation}
or
\begin{equation} \label{Greek singularities}
\kappa_{\pm}\,=\, {1\over 4\beta} \left\lbrace\, -(4+\det\AAA)
\,\pm\, \sqrt{(4-\det\AAA)^2+16|\gamma|^2} \right\rbrace\,.
\end{equation}
These singularities produce an eigenvalue provided the
corresponding root $\,\kappa\,$ is positive, otherwise we have a
zero-energy resonance (for $\,\kappa=0\,$) or an antibound state,
\ie, a resonance hidden deeply on the second sheet of the complex
energy surface, for $\,\kappa<0\,$.
\begin{proposition} \label{eigenvalues}
The operator $\,H_{\AAA}\,$ has at most two eigenvalues which are
given by the formula
\begin{equation} \label{lc eigenvalues}
\epsilon_{\pm}\,:=\, -\kappa_{\pm}^2\,=\, -\, {1\over 2}(a^2\!+b^2\!
+2|c|^2) \,\pm\, \sqrt{ {1\over 4}(a^2\!-b^2)^2+(a+b)^2|c|^2}
\end{equation}
provided the corresponding root $\,\kappa_{\pm}\,$ is positive, or
\begin{equation} \label{Greek eigenvalues}
\epsilon_{\pm}\,=\, -\, {8(2+|\gamma|^2)+(\det\AAA)^2\over 8\beta^2}
\,\pm\, {4+\det\AAA\over 16\beta^2}\, \sqrt{(4-\det\AAA)^2
+16|\gamma|^2}\,.
\end{equation}
The corresponding eigenfunction are $\,f_{\pm}:=f_{\kappa}\,$ for
$\,\kappa= \kappa_{\pm}\,$, where
\begin{equation} \label{eigenfunctions}
f_{\kappa}(x)\,:=\, \mu\Theta(x)\, e^{-\kappa x}+ \nu\Theta(-x)\,
e^{\kappa x}
\end{equation}
with the coefficients
\begin{equation} \label{coefficients}
\mu_{\pm}\,:=\, \sqrt{2\kappa_{\pm}(b+\kappa_{\pm}) \over
a+b+2\kappa_{\pm}} \,, \quad\;
\nu_{\pm}\,:=\,-\,{\bar c\over |c|}\, \sqrt{2\kappa_{\pm}
(a+\kappa_{\pm}) \over a+b+2\kappa_{\pm}} \,.
\end{equation}
\end{proposition}
\vspace{1mm}

\nid
{\em Proof:} The relations (\ref{lc eigenvalues}) and (\ref{Greek
eigenvalues}) are obtained by an elementary algebra. Since
$\,f_{\kappa}\,$ is the only square integrable solution to $\,f''\!+
\kappa^2 f=0\,$, it is sufficient to substitute it into the boundary
conditions to get (\ref{coefficients}). \quad \QED
\vspace{3mm}

Let us further mention some particular cases:
   \begin{description}
   \item{\em (i)} if $\,ab<0\,$, \ie, $\,4|\re\gamma|>|4+\det\AAA|\,$,
there is always one bound state and one antibound state,

   \item{\em (ii)} if both the decoupled operators on the halflines
refer to a repulsive interaction, $\,a,b>0\,$, it is still possible
to have a bound state, \ie, an eigenvalue of $\,H_{\AAA}\;$; this
{\em ``binding by conspiracy''} occurs \Iff $\;a\neq b\,$ and the
coupling is strong enough, $\,|c|> {1\over 2}\, \left\vert {a+b\over
a-b} \right\vert\,$,

   \item{\em (iii)} {\em two different eigenvalues} exist provided
both $\,a,b\,$ are negative, non-equal and the coupling of the
halflines is weak enough,
\begin{equation} \label{two bs}
|c|\,<\, {1\over 2}\, \left\vert {a+b\over a-b}\right\vert\,.
\end{equation}
In the parametrization (\ref{Greek bc}), these conditions acquire
a rather non-transparent form
\begin{eqnarray*}
(4+\det\AAA)\sgn\beta \!&>&\! \re\gamma \geq 0\,, \\
\beta^2(4+\det\AAA)^2 \!&>&\!
4|\re\gamma|^2 \left((4-\det\AAA)^2\!+16|\im\gamma|^2 \right)
\end{eqnarray*}

   \item{\em (iv)} the {\em eigenvalue crossing} occurs \Iff
$\;a=b\,$ and $\,c=0\,$. This, in turn, is clear also in the
parametrization of (\ref{Greek bc}): the condition $\,\det\AAA=4\,$
and $\,\gamma=0\,$ comprises of the decoupling requirement plus
$\,\re\gamma=0\,$, \ie, $\,a=b\,$,

   \item{\em (v)} the {\em $\,\delta$-interaction} does not fit well
into this scheme because of the lack of the parametrization (\ref{lc
bc}). Using the modification of Remark~\ref{HPA+uc}c, we can rewrite
the spectral condition as
$$
(1-ikA)(1-ikB)+k^2|C|^2=0\,.
$$
For $\,A=B=C=\alpha^{-1}\,$ it has the only solution $\,\kappa=
-\alpha/2\,$ which yields a bound state for $\,\alpha<0\,$. One can
also use the parametrization (\ref{Greek bc}): putting $\,\gamma=0\,$
we get
\begin{equation} \label{zero gamma}
\kappa_{\pm}\,:=\,\frac{-4-\alpha\beta\pm |4-\alpha\beta|}{4\beta}
\,=\, \left\lbrace \begin{array}{c} -\,{\alpha\over 2} \\ \\
-\,{2\over\beta} \end{array} \right.
\end{equation}
for all non-zero $\,\beta\,$, and only the upper solution survives
the limit $\,\beta\to 0\,$,

   \item{\em (vi)} the {\em $\,\delta'$-interaction}, on the other
hand, corresponds to $\,a=b=-c=\beta^{-1}\,$. The resolvent has again
a simple pole: the spectral condition $\,D_{\AAA}(k)=0\,$ is
solved by $\,\kappa=-2/\beta\,$ and $\,\kappa=0\,$, where the former
solution yields a bound state for $\,\beta<0\,$, while the latter
corresponds to no pole because of the vanishing residuum. We see also
that only the lower solution in (\ref{zero gamma}) is preserved in
the limit $\,\alpha=0\,$.
   \end{description}

\subsection{Scattering}

Using the observation made at the beginning of the preceding
subsection and the Birman-Kuroda theorem \cite[Thm.XI.9]{RS}, one
can check easily that the wave operators $\,\Omega_{\pm}(H_0,
H_{\AAA})\,$ exist and are asymptotically complete. It is also
straightforward to find that the on-shell scattering matrix is
\begin{equation} \label{S matrix}
S(k)\,=\, \left( \begin{array}{cc} t(k) & r(k) \\ -\bar r(k) & \bar
t(k) \end{array} \right)
\end{equation}
with
\begin{eqnarray*}
r(k) \!&=&\!-\, \frac{(a-ik)(b+ik)-|c|^2}{(a-ik)(b-ik)-|c|^2} \,=\,
2\,\frac{-\det\AAA+ (\gamma-ik\beta)(\bar\gamma-ik\beta)}
{(2-ik\beta)(\det\AAA-2ik\beta) -2|\gamma|^2}\,, \\ \\
t(k) \!&=&\! \frac{2ikc}{(a-ik)(b-ik)-|c|^2} \,=\,
-\,ik\beta\, \frac{4-\det\AAA +4i\im\gamma}
{(2-ik\beta)(\det\AAA-2ik\beta) -2|\gamma|^2}\,,
\end{eqnarray*}
and to check that it is unitary because $\,|r(k)|^2\!
+|t(k)|^2=1\,$. It follows from Proposition~\ref{bc} that there is
no transmission in the decoupled case, and we easily the standard
expressions corresponding to the particular cases of the
$\,\delta\,$ and $\,\delta'$-interactions
\cite[Chap.~I.3,4]{AGHH}.

The {\em low-} and {\em high-energy behaviour} of the GPI depends
substantially on the parameters. For small $\,k\,$ we have
\begin{eqnarray} \label{low energy}
r(k)\!&=&\! -1-\, {ik\over 2\alpha}(4+\det\AAA+4\re\gamma)
+\OO(k^2)\,, \nonumber \\ \\
t(k) \!&=&\!-\,{ik\over 2\alpha}(4-\det\AAA+4i\im\gamma)
+\OO(k^2) \nonumber
\end{eqnarray}
provided $\,\alpha\neq 0\;$; hence we have a full decoupling in
the limit $\,k\to 0\,$. On the other hand, if $\,\alpha=0\,$ we
find
$$ r(k)\,=\, {4\re\gamma-2ik\beta\over
4+|\gamma|^2-2ik\beta}\,, \quad\; t(k)\,=\,
{4-|\gamma|^2+4i\im\gamma\over 4+|\gamma|^2-2ik\beta}\,, $$
so the GPI is transparent in the low-energy limit \Iff
$\;\re\gamma=0\;$ (which includes the case of
$\,\delta'$-interaction) while in general both the reflection and
transmission amplitudes are non-zero.

At high energies the value of $\,\beta\,$ is important; if it is
non-zero then the $\,S$-matrix elements behave as
\begin{eqnarray} \label{high energy}
r(k)\!&=&\! -1+\,{i\over 2\beta k}(4+\det\AAA+4\re\gamma)
+\OO(k^{-2})\,, \nonumber \\ \\
t(k) \!&=&\! {i\over 2\beta k}(4-\det\AAA+4i\im\gamma)
+\OO(k^{-2})\,. \nonumber
\end{eqnarray}
Hence if the GPI contains a non-zero ``component'' of the
$\,\delta'$-interaction, it exhibits a full high-energy
decoupling. On the other hand, the limit $\,\beta\to 0\,$ yields
\begin{equation} \label{zero beta}
r(k)\,=\,-\, {2\alpha+ 4ik\re\gamma\over 2\alpha-
ik(4+|\gamma|^2)}\,, \quad\; t(k)\,=\,-ik\,
{4-|\gamma|^2+4i\im\gamma\over 2\alpha- ik(4+|\gamma|^2)}\,,
\end{equation}
so the GPI is transparent in the high-energy limit \Iff
$\;\re\gamma=0\;$ (which includes the case of
$\,\delta$-interaction) while in general again neither the
reflection nor transmission are suppressed. Notice the remarkable
duality between the scattering properties at low and high energies
when the roles of $\,\alpha\,$ and $\,\beta\,$ are switched.

\setcounter{equation}{0}
\section{The geometric phase}

Let us investigate the phase resulting from a parameter change.
For simplicity, consider the case $\,a=b\,$ with
$\,c=|c|\,e^{i\xi}\;$; this corresponds to
$$ \AAA\,=\, {2\over a-|c|\cos\xi}\: \left(
\begin{array}{cc} a^2\!-|c|^2 & -i|c|\,\sin\xi \\ -i|c|\,\sin\xi &
1 \end{array} \right)\,. $$
Then we have $\,\kappa_{\pm}=-a\mp|c|\,$ and the coefficients
(\ref{coefficients}) are $\,\mu_{\pm}= \sqrt{-a\mp|c|}\,$ and
$\,\nu_{\pm}= -\,e^{-i\xi}\sqrt{-a\mp|c|}\,$ so
\begin{eqnarray*}
\lefteqn{df_{\kappa_{\pm}}(x)= \Bigl\lbrace\, {1\over 2(|c|\pm a)}\,
f_{\kappa_{\pm}}(x)\,\pm\, x\Bigl\lbrack \sqrt{-a\mp|c|}\,
\Theta(x)\, e^{(a\pm|c|)x} } \\
&& +\, e^{-i\xi} \sqrt{-a\mp|c|}\,\Theta(-x)\, e^{-(a\pm|c|)x}
\Bigr\rbrack \Bigr\rbrace\, d|c| +\,i\, e^{-i\xi}
\sqrt{-a\mp|c|}\,\Theta(-x)\, e^{-(a\pm|c|)x}\, d\xi \,.
\end{eqnarray*}
As a simple example, consider $\,|c|\,$ fixed and let $\,\xi\,$
run through $\,[0,2\pi)\,$, then we obtain a non-trivial Berry
phase,
$$ \int_0^{2\pi}\, i(f_{\kappa_{\pm}},df_{\kappa_{\pm}})\,=\,
\int_0^{2\pi} d\xi\, (-a\mp|c|)\, \int_{-\infty}^0\,
e^{-2(a\pm|c|)x} dx \,=\, {1\over 2}\, \int_0^{2\pi} d\xi \,=\,
\pi\,, $$
independently of $\,|c|\,$.

\setcounter{equation}{0}
\section{Arrays of generalized point interactions}

Consider an equidistant array of GPI's supported by the lattice
$\,\LL:= \{n\ell\}_{n=-\infty}^{\infty}\,$ with a spacing
$\,\ell>0\,$. Let the boundary conditions at the $\,n$-th lattice
point be given by $\,\AAA_n\;$; for simplicity, we shall restrict
our attention to the case when none of them is separating, \ie,
$\,\det\AAA_n\neq 4\,$ or $\,\im\gamma_n\neq 0\,$ holds for each
$\,n\,$.

We denote the corresponding operator by $\,H(\{\AAA_n\},\LL)\;$; it
acts as the free Hamiltonian outside $\,\LL\,$, \ie,
$\,(H(\{\AAA_n\},\LL)f)(x)= -f''(x)\,$ for $\,n\ell<x<
(n\!+\!1)\ell\,$ and at the points $\,x=n\ell\,$ the functions of
$\,D(H(\{\AAA_n\},\LL))\,$ satisfy the boundary conditions of the
form (\ref{Greek bc}) with the coefficients given by $\,\AAA_n\,$.

In particular, if all the $\,\AAA_n\,$ are the same, $\,\AAA_n=
\AAA\,$, we write $\,H(\{\AAA_n\},\LL)=: H(\AAA,\ell)\,$. This
corresponds to a periodic system and one expects it to have a
band-type spectrum.

\begin{theorem} \label{KP}
The spectrum of $\,H(\AAA,\ell)\,$ with a non-separating
$\,\AAA\neq 0\,$ is purely absolutely continuous and of the form
$\,\sigma(H(\AAA,\ell)) =\bigcup_{m=0}^{\infty}
\Delta_m(\AAA,\ell)\,$, where $\,\Delta_m(\AAA,\ell)\,$ are
mutually disjoint closed intervals, the lowest of which may be
empty.
   \begin{description}
   \item{(a)} If $\,\beta\neq 0\,$, the spectral bands
$\,\Delta_m(\AAA,\ell)\,$ are centered roughly at the values
\begin{equation} \label{band centre}
\epsilon_m \,:= \, \pi^2 m^2\,+\,(-1)^m\,
{2(4+\det\AAA)\over \beta\ell}\,+\, \OO(m^{-1})
\end{equation}
and their widths are asymptotically constant at high energies,
\begin{equation} \label{band width 1}
|\Delta_m(\AAA,\ell)|\,=\,\frac{2\sqrt{(4-\det\AAA)^2\!+16|\im\gamma|^2}}
{|\beta|\ell}\,+\, \OO(m^{-1})\,.
\end{equation}
It follows that the width $\,|\Gamma_m(\AAA,\ell)|\,$ of the
$\,m$-th gap is growing linearly up to higher-order terms as
$\,m\to\infty\,$.
   \item{(b)} If $\,\beta=0\,$ and $\,\re\gamma\neq 0\,$, the widths
of both bands and gaps are growing,
\begin{equation} \label{band width 2}
|\Delta_m(\AAA,\ell)|\,=\,{4\pi m\over\ell}\, \arcsin\left(
\frac{\sqrt{(4-|\gamma|^2)^2\!+16|\im\gamma|^2}}
{4+|\gamma|^2}\right) \left(\,1+ \OO(m^{-1}) \right)\,,
\end{equation}
\begin{equation} \label{gap width 1}
|\Gamma_m(\AAA,\ell)|\,=\,{4\pi m\over\ell}\, \arccos\left(
\frac{\sqrt{(4-|\gamma|^2)^2\!+16|\im\gamma|^2}}
{4+|\gamma|^2}\right) \left(\,1+ \OO(m^{-1}) \right)\,.
\end{equation}

   \item{(c)} If $\,\beta=0\,$ and $\,\re\gamma=0\,$, the $\,m$-th
gap has $\,\pi^2 m^2\,$ as one endpoint and its width is asymptotically
constant,
\begin{equation} \label{gap width 2}
|\Gamma_m(\AAA,\ell)|\,=\,{8|\alpha|\over (4+|\gamma|^2)\ell}\,+\,
\OO(m^{-1})\,.
\end{equation}
Consequently, the band widths $\,|\Delta_m(\AAA,\ell)|\,$ grow
linearly up to higher-order terms as $\,m\to\infty\,$.
   \end{description}
\end{theorem}
\vspace{1mm}

\nid
{\em Proof:} Following the standard Bloch decomposition we have to
find eigenvalues of the GPI Hamiltonian on $\,L^2(-\ell/2,\ell/2)\,$
with the boundary conditions
\begin{equation} \label{Bloch}
f\left(-\,{\ell\over 2} \right)\,=\, e^{i\theta} f\left({\ell\over 2}
\right)\,, \quad\; f'\left(-\,{\ell\over 2} \right)\,=\, e^{i\theta}
f'\left({\ell\over 2} \right)\,.
\end{equation}
In combination with (\ref{Greek bc}), it requires the determinant
$$
\left\vert \begin{array}{cccc} {\alpha\over 2}\,+ik\left(1
+\,{\gamma\over 2}\right) & {\alpha\over 2}\,-ik\left(1
+\,{\gamma\over 2}\right) & {\alpha\over 2}\,-ik\left(1
-\,{\gamma\over 2}\right) & {\alpha\over 2}\,+ik\left(1
-\,{\gamma\over 2}\right) \\ \\ 1-\, {\bar\gamma\over 2}\,+ ik\beta &
1-\, {\bar\gamma\over 2}\,- ik\beta & -1-\, {\bar\gamma\over 2}\,+
ik\beta & -1-\, {\bar\gamma\over 2}\,-ik\beta \\ \\ e^{-ik\ell/2} &
e^{ik\ell/2} & -e^{i(\theta+k\ell/2)} & -e^{i(\theta-k\ell/2)} \\ \\
e^{-ik\ell/2} & -e^{ik\ell/2} & -e^{i(\theta+k\ell/2)} &
e^{i(\theta-k\ell/2)} \end{array} \right\vert
$$
to be zero, which yields the band condition
\begin{equation} \label{band condition 1}
\re\left((4-\det\AAA+i\im\gamma)\,e^{i\theta} \right)\,=\,
(4+\det\AAA) \cos k\ell\,+\, {2\over k}(\alpha-\beta k^2) \sin
k\ell\,.
\end{equation}
For $\,\beta\neq 0\,$ the \rhs is asymptotically dominated by growing
oscillations coming from the last term; finding its zeros and
expanding around them we prove the assertion~{\em (a)}. If
$\,\beta=0\,$, we can rewrite the band condition as
\begin{equation} \label{band condition 2}
\re\left(t(\infty)\,e^{i\theta} \right)\,=\, \cos k\ell\,+\,
{2\alpha\over k(4+|\gamma|^2)} \sin k\ell\,,
\end{equation}
where $\,t(\infty):= \lim_{k\to\infty} t(k)\,$ is given by
(\ref{zero beta}). Suppose first that $\,\re\gamma\neq 0\;$ (and
$\,\gamma\neq \pm 2\,$ because the GPI is non-separating by
assumption), then $\,0< |t(\infty)| <1\,$. The \rhs is
asymptotically dominated by the first term; this yields {\em (b)}.
Finally, for $\,\re\gamma=0\,$ we can adapt easily the standard
Kronig-Penney argument \cite[Chap.III.2]{AGHH} with $\,\alpha\,$
replaced by $\,\alpha(4+|\gamma|^2)^{-1}$. \quad \QED \vspace{3mm}

With the stated motivation in mind, we have concentrated on the
infinite number of gaps and their asymptotic behaviour, using a
not fully standard band numbering. We shall not discuss other
properties such as the bottom of the spectrum, band profiles {\em
etc.}; they can be obtained in the same way as in the particular
cases of the $\,\delta\,$ and $\,\delta'$-interactions -- \cf
\cite[Chaps.~III.2,3]{AGHH}.

The main conclusion of the theorem is that the high-energy
behaviour of the generalized Kronig-Penney model reflects that of
the one-center GPI. If there is a non-vanishing ``component'' of
the $\,\delta'$-interaction in $\,H_{\AAA}\,$ leading to the
high-energy decoupling, the corresponding $\,H(\AAA,\ell)\,$ has
the gap-to-band width ratio growing approximately linearly with
the band number.

On the other hand, the case {\em (c)} exhibits the $\,\delta$-type
behaviour with widening bands and asymptotically constant gaps. A
new type of behaviour intermediate between the $\,\delta\,$ and
$\,\delta'\,$ extremes corresponds to the case {\em (b)}: here
both gaps and bands are widening and the ratio of their width is
asymptotically constant.

\section*{Acknowledgement}
P.E. thanks the E.~Schr\"odinger Institute where this work was
done for the hospitality extended to him. The research was
partially supported by the GAAS Grant No.~14814.


\begin{thebibliography}{article}

\bibitem{ABD}
S.~Albeverio, Z.~Brzezniak, L.~Dabrowski: Fundamental solutions of
the heat and Schr\"odinger equations with point interaction, {\em
J.~Funct.~Anal.} {\bf 130} (1995), 220--254.
   \vspace{-1.8ex}
\bibitem{AGHH}
S.~Albeverio, F.~Gesztesy, R.~H\o egh-Krohn, H.~Holden: {\em
Solvable Models in Quantum Mechanics}, Springer, Heidelberg 1988.
   \vspace{-1.8ex}
\bibitem{AGH}
S.~Albeverio, F.~Gesztesy, H.~Holden: Comments on a recent note on
the Schr\"odinger equation with a $\,\delta'$--interaction, {\em
J.~Phys.} {\bf A26} (1993), 3903--3904.
   \vspace{-1.8ex}
\bibitem{AESS}
J.-P.~Antoine, P.~Exner, P.~\v Seba, J.~Shabani: A mathematical
model of heavy--quarkonia decay, Ann.~ Phys. {\bf 233} (1994),
1--16.
   \vspace{-1.8ex}
\bibitem{AEL1}
J.E.~Avron, P.~Exner, Y.~Last: {\em Periodic Schr\"odinger
operators with large gaps and Wannier--Stark ladders}, Phys. Rev.
Lett. {\bf 72} (1994), 896--899.
   \vspace{-1.8ex}
\bibitem{E2}
P.~Exner: {\em The absence of the absolutely continuous spectrum
for $\,\delta'$ Wannier--Stark ladders}, J. Math. Phys.  {\bf 36}
(1995), 4561--4570.
   \vspace{-1.8ex}
\bibitem{BF}
F.A.~Berezin, L.D.~Faddeev : A remark on Schr\"odinger equation
with a singular potential, {\em Sov.~Acad.~Sci.~Doklady} {\bf 137}
(1961), 1011-1014 (in Russian).
   \vspace{-1.8ex}
\bibitem{Ca}
M.~Carreau: Four--parameter point--interactions in 1D quantum
systems, J.~Phys. {\bf A26} (1993), 427--432.
   \vspace{-1.8ex}
\bibitem{CH}
P.R.~Chernoff, R.~Hughes: A new class of point interactions in one
dimension {\em J.~Funct.~Anal.} {\bf 111} (1993), 92--117.
   \vspace{-1.8ex}
\bibitem{E1}
P.~Exner: A solvable model of two--channel scattering, {\em
Helv.~Phys.~Acta} {\bf 64} (1991), 592--609.
   \vspace{-1.8ex}
\bibitem{Fe}
E.~Fermi: Sul moto dei neutroni nelle sostanze idrogenate, {\em
Ricerca Scientifica} {\bf 7} (1936), 13--52 (English translation
in {\em E.~Fermi Collected Papers, Vol.~I, Italy, 1921--1938},
University of Chicago Press 1962; pp.~980--1016.
   \vspace{-1.8ex}
\bibitem{GeK}
F.~Gesztesy, W.~Kirsch: One--dimensional Schr\"odinger operators
with interactions singular on a discrete set, {\em J.~Reine
Angew.~Math.} {\bf 362} (1985), 28--50.
   \vspace{-1.8ex}
\bibitem{Gr}
D.J.~Griffith: Boundary conditions at the derivative of a delta
function, {\em J.~Phys.} {\bf A26} (1993), 2265--2267.
\vspace{-1.8ex}
\bibitem{GK}
H.~Grosse, W.L.~Kennedy: The geometric phase in a simple model,
{\em Phys.~ Lett.} {\bf A154} (1991), 116--122.
   \vspace{-1.8ex}
\bibitem{KP}
R.~de L.~Kronig, W.G.~Penney: Quantum mechanics of electrons in
crystal lattices, {\em Proc.~Roy.~Soc. (London)} {\bf 130A}
(1931), 499--513.
   \vspace{-1.8ex}
\bibitem{Ku}
P.~Kurasov: {\em On direct and inverse scattering problems in
dimension one}, PhD Thesis, Stockholm University 1993.
   \vspace{-1.8ex}
\bibitem{Ne}
G.~Nenciu: Dynamics of band electrons in electric and magnetic
fields: rigo\-rous justification of the effective Hamiltonians,
{\em Rev.~Mod.~Phys.} {\bf 63} (1993), 91--127. \vspace{-1.8ex}
\bibitem{RS}
M.~Reed, B.~Simon: {\em Methods of Modern Mathematical Physics,
III.~Scattering Theory, IV.~Analysis of Operators}, Academic
Press, New York 1979, 1978.
   \vspace{-1.8ex}
\bibitem{Se1}
P.~\v Seba: The generalized point interaction in one dimension,
{\em Czech.~J.~Phys.} {\bf B36} (1986), 667--673.
   \vspace{-1.8ex}
\bibitem{Se2}
P.~\v Seba: Some remarks on the $\,\delta'$--interaction in one
dimension, {\em Rep. Math. Phys.} {\bf 24} (1986), 111--120.
   \vspace{-1.8ex}
\begin{center}
* \quad * \quad *
\end{center}

\bibitem{ADK}
S.~Albeverio, L.~Dabrowski, P.~Kurasov: Symmetries of
Schr\"odinger operators with point interactions, {\em Lett. Math.
Phys.} {\bf 45} (1998), 33--47.
   \vspace{-1.8ex}
\bibitem{AK}
S.~Albeverio, P.~Kurasov: {\em Singular Perturbations of Differential
Operators}, Cambridge  University Press 1999.
   \vspace{-1.8ex}
\bibitem{ADE}
J.~Asch, P.~Duclos, P.~Exner: Stability of driven systems with
growing gaps. Quantum rings and Wannier ladders, {\em J. Stat.
Phys.} {\bf 92} (1998), 1053--1069.
   \vspace{-1.8ex}
   \bibitem{AB}
J.E. Avron, J. Berger: Adiabatic response of quantum systems
pinching a gap closure, {\em Chem. Phys. Lett.} {\bf 294} (1998),
13--18.
   \vspace{-1.8ex}
   \bibitem{CS}
T.~Cheon, T.~Shigehara: Realizing discontinuous wave functions
with renormalized short-range potentials , {\em Phys. Lett.} {\bf
A243} (1998), 111--116.
   \vspace{-1.8ex}
   \bibitem{ETV}
P.~Exner, M.~Tater, D.~Van\v{e}k: A single-mode quantum transport
in serial-structure geometric scatterers, {\em in preparation}
   \vspace{-1.8ex}
   \bibitem{FT}
T.~F\"ul\"op, I.~Tsutsui: A free particle on a circle with point,
{\em quant-ph/9910062}
   \vspace{-1.8ex}
\bibitem{Kis}
A.~Kiselev: Some examples in one--dimensional ``geometric"
scattering on manifolds, {\em J. Math. Anal. Appl.} {\bf 212}
(1997), 263--280.
   \vspace{-1.8ex}
\bibitem{RT}
J.M.~Roman, R.~Tarrach: The regulated one-dimensional point
interactions, {\em J. Phys.} {\bf A29} (1996), 6073--6085.
   \vspace{-1.8ex}
   \bibitem{SA}
L. Sadun, J.E. Avron: Adiabatic curvature and the S-matrix, {\em
Commun. Math. Phys.} {\bf 181} (1996), 685--710.
   \vspace{-1.8ex}

   \end{thebibliography}
\end{document}